\documentstyle[12pt,aaspp4]{article}
\begin{document}

\title{Extracting Energy from a Black Hole through Its Disk}

\author{Li-Xin Li}
\affil{Princeton University Observatory, Princeton, NJ 08544--1001, USA}
\affil{E-mail: lxl@astro.princeton.edu}

\begin{abstract}
When some magnetic field lines connect a Kerr black hole with a disk 
rotating around it, energy and angular momentum are transferred between them. 
If the black hole rotates faster than the disk, $ca/GM_H>0.36$ for a thin 
Keplerian disk, then energy and angular momentum are extracted from the black 
hole and transferred to the disk ($M_H$ is the mass and $a M_H$ is the 
angular momentum of the black hole).  This way the energy originating
in the black hole may be radiated away by the disk.

The total amount of energy that can be extracted from the black hole
spun down from $ca/GM_H = 0.998$ to $ca/GM_H = 0.36$ by a thin Keplerian
disk is $\approx 0.15 M_Hc^2$.  This is larger than $\approx 0.09 M_Hc^2$ 
which can be extracted by the Blandford-Znajek mechanism.
\end{abstract}


\section{Introduction}
Extraction of energy from a black hole or an accretion disk through
magnetic braking has been investigated by many people. As a rotating black hole
is threaded by magnetic field lines which connect with remote astrophysical loads, 
energy and angular momentum
are extracted from the black hole and transported to the remote loads via Poynting
flux (Blandford \& Znajek 1977; Macdonald \& Thorne 1982; Phinney 1983). This is
usually called the Blandford-Znajek mechanism and has been suggested to be
a plausible process for powering jets in active galactic nuclei (Rees, Begelman,
Blandford, \& Phinney 1982; Begelman, Blandford, \& Rees 1984) and gamma ray
bursts (Paczy\'nski 1993; Lee, Wijers, \& Brown 1999). Similar
process can happen to an accretion disk when some of magnetic field
lines threading the disk are open and connect with remote astrophysical loads
(Blandford 1976; Blandford \& Znajek 1977; Macdonald \& Thorne 1982; Livio,
Ogilvie, \& Pringle 1999; Li 1999).

In this paper we investigate the effects of magnetic field lines connecting a
Kerr black hole with a disk surrounding it.  This kind of
magnetic field lines are expected to exist and have important effects (Macdonald
\& Thorne 1982; Blandford 1999, 2000; Gruzinov 1999).  We find that, with the
existence of such magnetic coupling between the black hole and the disk, energy
and angular momentum are transfered between them. If the black hole rotates faster 
than the disk, energy and angular momentum are extracted from the black hole and 
transferred to the disk via Poynting flux. This is the case when
$a/M_H>0.36$ for a thin Keplerian  disk, where $M_H$ is the mass of the 
black hole and $aM_H$ is the angular momentum of the black hole. 
Throughout the paper we use the geometric units with $G = c = 1$.  The energy 
deposited into the disk by the black hole is eventually radiated to infinity
by the disk. This provides a way for extracting energy from a black hole through its
disk. If the disk has no accretion (or the accretion rate is very low), 
the power of the disk 
essentially comes from the rotational energy of the black hole. We will show
that the magnetic coupling between the black hole and the disk has a higher
efficiency in extracting energy from a Kerr black hole than the Blandford-Znajek
mechanism.

\section{Transfer of Energy and Angular Momentum between a Black Hole and Its
Disk by Magnetic Coupling}
Suppose a bunch of magnetic field lines connect a rotating black
hole with a disk surrounding it. Due to the rotation of the black hole
and the disk, electromotive forces are induced on both the black hole's horizon 
and the disk (Macdonald \& Thorne 1982; Li 1999)
\begin{eqnarray}
    {\cal E}_H = {1\over 2\pi}\Omega_H \Delta\Psi\,,  \hspace{1cm}
    {\cal E}_D = -{1\over 2\pi}\Omega_D \Delta\Psi\,,
    \label{emf}
\end{eqnarray}
where $\Omega_H$ is the angular velocity of the black hole, $\Omega_D$ is the
angular velocity of the disk, $\Delta\Psi$ is the magnetic flux connecting
the black hole with the disk. The black hole and the disk form a closed
electric circuit, the electric current flows through the magnetic field lines 
connecting them. Suppose the disk and the black hole rotates in the same direction, 
then ${\cal E}_H$ and ${\cal E}_D$ have opposite signs. This means that energy
and angular momentum are transferred
either from the black hole to the disk or from the disk to the black hole, the
direction of transfer is determined by the sign of ${\cal E}_H + {\cal E}_D$. By the
Ohm's law, the current is $I = ({\cal E}_H+{\cal E}_D)/Z_H =
\Delta\Psi(\Omega_H-\Omega_D)/(2\pi Z_H)$, where $Z_H$ is the resistance of the 
black hole which is of several hundred Ohms (the disk is perfectly conducting so
its resistance is zero). The power deposited into the disk by the black hole is
\begin{eqnarray}
    P_{HD} = - I {\cal E}_D
            = \left({\Delta\Psi\over 2\pi}\right)^2 \,{\Omega_D
            \left(\Omega_H - \Omega_D\right)
            \over Z_H}\,.
    \label{pow3}
\end{eqnarray}
The torque on the disk produced by the black hole is
\begin{eqnarray}
    T_{HD} = {I\over 2\pi}\Delta\Psi =
             \left({\Delta\Psi\over 2\pi}\right)^2 \,{\left(\Omega_H -
	     \Omega_D\right)
             \over Z_H}\,.
    \label{toq}
\end{eqnarray}
As expected, we have $P_{BH} = T_{BH}\Omega_D$.

The signs of $P_{HD}$ and $T_{HD}$ are determined by the sign of 
$\Omega_H-\Omega_D$.  When $\Omega_H > \Omega_D$, we have $P_{HD}>0$ 
and $T_{HD}>0$, energy and angular
momentum are transferred from the black hole to the disk.
When $\Omega_H < \Omega_D$, we have $P_{HD} < 0$ and $T_{HD}<0$, energy 
and angular momentum are transferred from the disk to the black hole so the
black hole is spun up. For a disk with non-rigid rotation, $\Omega_D$
varies with radius. For fixed values of $\Delta\Psi$,
$\Omega_H$, and $Z_H$, $P_{HD}$ peaks at $\Omega_D = \Omega_H/2$.
However for realistic cases which is most important is when the
magnetic field lines touch the disk close to the inner boundary,
so $\Omega_D$ in Eq.~(\ref{pow3}) and Eq.~(\ref{toq}) can be taken 
to be the value
at the inner boundary of the disk. According to Gruzinov (1999) the
magnetic fields will be more unstable against screw instability if the 
foot-points 
of the field lines on the disk are far from the inner boundary of the disk.

For a thin Keplerian disk around a Kerr black hole in the equatorial plane, the
angular velocity of the disk is (Novikov \& Thorne 1973)
\begin{eqnarray}
    \Omega_D(r)=\left({M_H\over r^3}\right)^{1/2}{1\over 1+a\left(M_H/r^3
             \right)^{1/2}}\,,
    \label{wd}
\end{eqnarray}
where $r$ is the Boyer-Lindquist radius in Kerr spacetime. $\Omega_D(r)$
decreases with increasing $r$. The angular velocity of a Kerr black hole is
\begin{eqnarray}
    \Omega_H = {a\over 2M_H r_H}\,,
    \label{wh}
\end{eqnarray}
where $r_H = M_H + \sqrt{M_H^2-a^2}$ is the radius of the event horizon.
$\Omega_H$ is constant on the horizon. The inner boundary of a Keplerian disk
is usually assumed to be at the marginally stable orbit with radius
(Novikov \& Thorne 1973)
\begin{eqnarray}
    r_{ms}=M_H\left\{3+z_2-\left[(3-z_1)(3+z_1+2z_2)\right]^{1/2}\right\}\,,
    \label{rms}
\end{eqnarray}
where
\begin{eqnarray}
    z_1=1+\left(1-a^2/M_H^2\right)^{1/3}\left[\left(1+a/M_H\right)^{1/3}+
    \left(1-a/M_H\right)^{1/3}\right]\,,
    \label{rms2}
\end{eqnarray}
and
\begin{eqnarray}
    z_2=\left(3a^2/M_H^2+z_1^2\right)^{1/2}\,.
    \label{rms3}
\end{eqnarray}
Inserting Eq.~(\ref{rms}) into
Eq.~(\ref{wd}), we obtain the angular velocity of the disk at its inner 
boundary:
$\Omega_{ms} = \Omega_D(r_{ms})$. For the Schwarzschild case (i.e $a = 0$) we 
have $r_{ms} = 6 M_H$ and $\Omega_{ms} = 6^{-3/2}M_H^{-1}\equiv\Omega_0$.

Assuming the magnetic field lines touch the disk close to the inner boundary, 
we have $P_{HD} \approx P_0 f$
where
\begin{eqnarray}
    P_0 = \left({\Delta\Psi\over 2\pi}\right)^2 {\Omega_0^2\over Z_H}
    \label{psch}
\end{eqnarray}
is the value of $-P_{HD}$ for the Schwarzschild case, and
\begin{eqnarray}
    f = {\Omega_{ms}\left(\Omega_H -\Omega_{ms}\right)\over \Omega_0^2}
    \label{ratio}
\end{eqnarray}
is a function of $a/M_H$ only. The variation of $P_{HD}$ with $a/M_H$
is shown in Fig.~\ref{figure1}. We see that $P_{HD}>0$ when $0.36 < a/M_H <1$,
$P_{HD}<0$ when $0\le a/M_H<0.36$. $P_{HD}=0$ at $a/M_H \approx 0.36$ and
$a/M_H=1$ since $P_{HD}\propto \Omega_H-\Omega_{ms}$ and $\Omega_H = \Omega_{ms}$ 
when $a/M_H\approx 0.36$ and $a/M_H =1$.
For fixed $\Delta\Psi$, $M_H$, and $Z_H$, $P_{HD}$ peaks
at $a/M_H\approx 0.981$.  $T_{HD}$ always has the same sign as $P_{HD}$ since
$P_{HD} = T_{HD}\Omega_D$ for a perfectly conducting disk.

\section{Extracting Energy from a Black Hole through Its Disk}
When $a/M_H>0.36$, energy and angular momentum are extracted from the black hole
and transferred to the disk. So a fast rotating black hole can pump its rotational
energy into a disk surrounding it through magnetic coupling between them.
Once the energy gets into the disk, it can be radiated to infinity either
in the form of Poynting flux associated with jets or winds, or in the form
of thermal radiation associated with dissipative processes in the disk.
If the disk is not accreting or its accretion rate is very low,
then the disk's power 
comes from the rotational energy of the black hole.
This provides a way for {\em indirectly} extracting energy from a rotating 
black hole.  Note, that the Blandford-Znajek mechanism is a way for 
{\em directly} extracting energy from a rotating black hole
to the remote load.

It is possible that the Blandford-Znajek mechanism provides a very ``clean''
energy beam, while energy extracted from the disk is ``dirty'', contaminated
by matter from the disk corona (R. D. Blandford 1999a, private communication).
However, we must keep in mind that there exists no quantitative model 
demonstrating how to generate clean energy with the Blandford-Znajek process.

Let us consider again our case, in which
Kerr black hole loses its energy and angular momentum through the 
magnetic interaction with a thin Keplerian disk, with the magnetic field lines 
touching the disk close to the marginally stable orbit. 
The evolution of the mass 
and angular momentum of the black hole are given by
\begin{eqnarray}
    {d M_H\over dt} = -2 P_{HD}\,,
    \hspace{1cm} {d J_H\over dt} = -2 T_{HD}\,,
    \label{evol}
\end{eqnarray}
where $P_{HD}$ and $T_{HD}$ are given by Eq.~(\ref{pow3}) and Eq.~(\ref{toq})
respectively, the factors $2$ come from the fact that a disk has
two faces. From Eq.~(\ref{evol}) we obtain ${dJ_H/ dM_H} = {1/\Omega_{ms}}$,
where we have used $P_{HD} \approx T_{HD}\Omega_{ms}$. Define the spin of a Kerr
black hole by $s \equiv a/M_H = J_H/M_H^2$, then we have
\begin{eqnarray}
    {ds\over d\ln M_H} = {1\over \omega} - 2 s\,,
    \label{dsm}
\end{eqnarray}
where $\omega \equiv M_H\Omega_{ms}$ is a function of $s$ only. Eq.~(\ref{dsm})
can be integrated
\begin{eqnarray}
    M_H(s) = M_{H,0} \exp\int_{s_0}^s{ds\over\omega^{-1}-2s}\,,
    \label{mh}
\end{eqnarray}
where $M_{H,0} = M_H(s=s_0)$. Consider a Kerr black hole with initial mass $M_H$ 
and the initial spin $s = 0.998$, which is
the maximum value of $s$ that an astrophysical black hole can have (Thorne
1974). As the black hole spins down to $s = 0.36$, the total amount of energy
extracted from the black hole by the disk can be calculated with Eq.~(\ref{mh}):
$\Delta E\approx 0.15 M_H$.
This amount of energy will eventually be transported to infinity by the disk.
In a realistic case the magnetic field lines touch the disk not exactly at the 
marginally stable orbit, the averaged angular velocity of the disk will be 
somewhat smaller than $\Omega_{ms}$, then the total amount of energy that 
can be extracted
from the black hole should be somewhat smaller than $0.15 M_H$.

For comparison let's calculate the amount of energy that can be extracted 
from a Kerr black hole by the Blandford-Znajek mechanism in the optimal case 
i.e. when the impedance matching condition is satisfied (cf. Macdonald \&
Thorne 1982). To do so, we only need to replace $\Omega_{ms}$ with
$\Omega_H/2$ in Eq.~(\ref{mh}), since the power and torque of the black hole
are related by $P_H = T_H\Omega_F$ where $\Omega_F$ is the angular velocity of 
magnetic field lines, and in the optimal case $\Omega_F = \Omega_H/2$.
Then we obtain that as the black hole spins down from $s = 0.998$ to $s = 0$ the
total energy extracted from the black hole by the Blandford-Znajek mechanism
is $\approx 0.09 M_H$.

We find that the magnetic coupling between a black hole and a disk has a higher
efficiency in extracting energy from the black hole than the Blandford-Znajek 
mechanism (see Fig.~\ref{figure2}). This is because the energy extracted 
from the black hole by the magnetic coupling to the disk has a 
larger ratio of energy to angular momentum than is the case for
the Blandford-Znajek mechanism.

\section{Conclusions}
When a black hole rotates faster than the disk,
which is the case if $a/M_H>0.36$ for a Kerr black hole with a 
thin Keplerian disk, then the black hole exerts a torque at the
inner edge of the disk. The torque transfers energy
and angular momentum from the black hole to the disk.
This is similar to the ``propeller'' mechanism in the case of a 
magnetized neutron star with a disk (Illarionov \& Sunyaev 1975). 
The energy transfered to the disk 
is eventually radiated to infinity by the disk. This provides a mechanism for
extracting energy from a black hole through its disk. 
For a Kerr black hole with the initial mass $M_H$ and spin $a/M_H = 0.998$, 
the total amount of energy that can be extracted by a thin Keplerian
disk is $\approx 0.15 M_H$.
Therefore, this is more efficient than the Blandford-Znajek mechanism
which can extract only $\approx 0.09 M_H$.

When the black hole rotates slower than the disk, i.e. $0\le a/M_H<0.36$,
energy and angular momentum are transferred from the disk to the black hole,
and the disk accretes onto the black hole.

\acknowledgments{I am very grateful to Bohdan Paczy\'nski for encouraging and 
stimulating discussions. This work was supported by the NASA grant NAG5-7016.}


\newpage
\figcaption[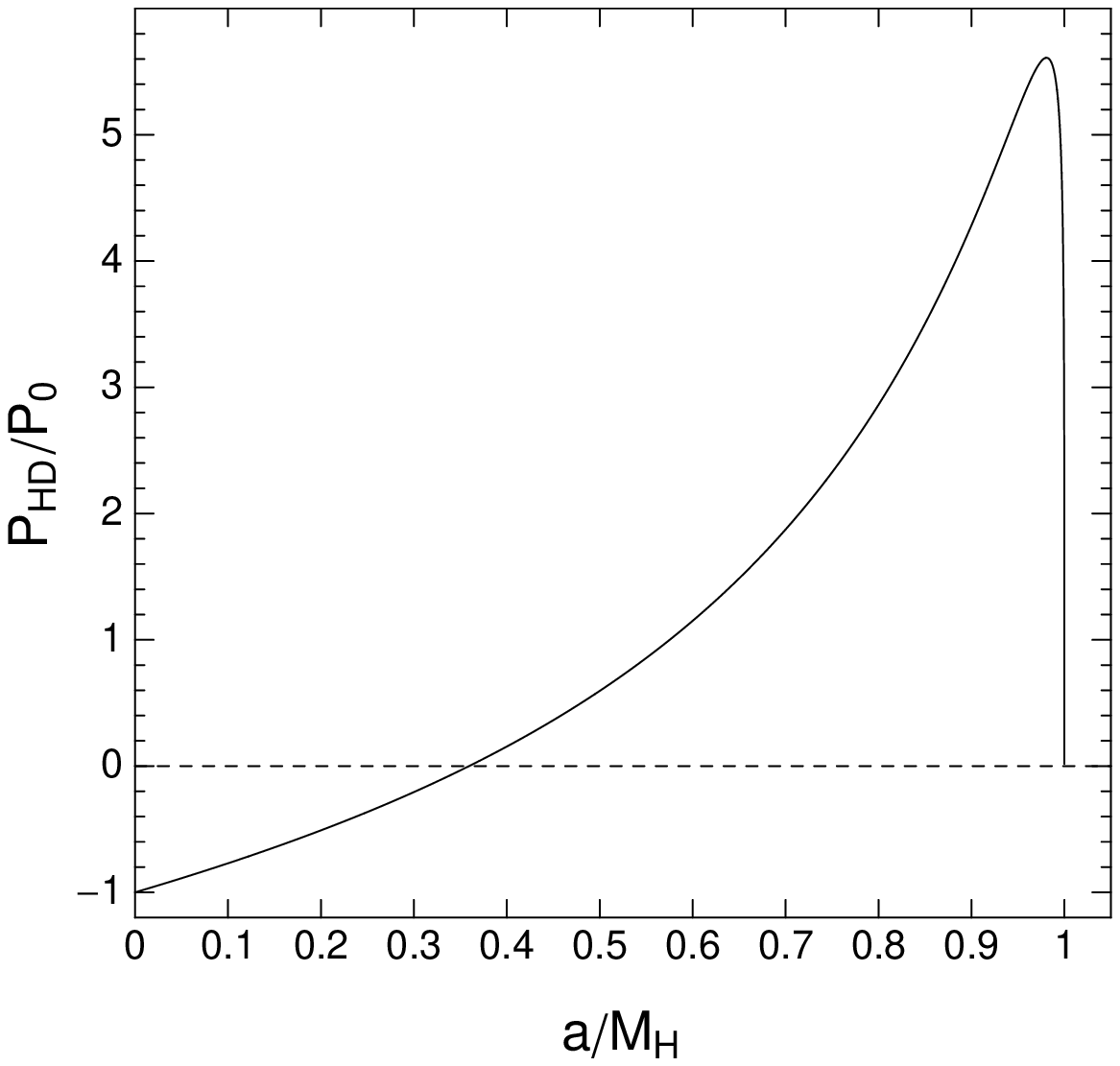]{Magnetic field lines connecting a black hole with an accretion
disk can transfer energy and angular momentum between them. In the figure is shown 
the dependence of the power of the energy transfer on the spin of the black hole
for the model of a Kerr black hole with a thin Keplerian disk. The magnetic field lines
are assumed to touch the disk close to the marginally stable orbit. The vertical 
axis shows the power $P_{HD}$ in unit of $P_0$, where $P_0$ is the value
of $-P_{HD}$ for the Schwarzschild case. The horizontal axis shows $a/M_H$, where
$M_H$ is the mass of the black hole, $a M_H$ is the angular momentum of the black
hole. If $P_{HD}>0$, which is the case when $0.36<a/M_H <1$, energy and angular 
momentum are transferred from the black hole to the disk; if $P_{HD}<0$, which 
is the case when $0\le a/M_H <0.36$, energy and angular momentum are transferred from
the disk to the black hole. $P_{HD}$ peaks at $a/M_H\approx 0.981$.
\label{figure1}}

\figcaption[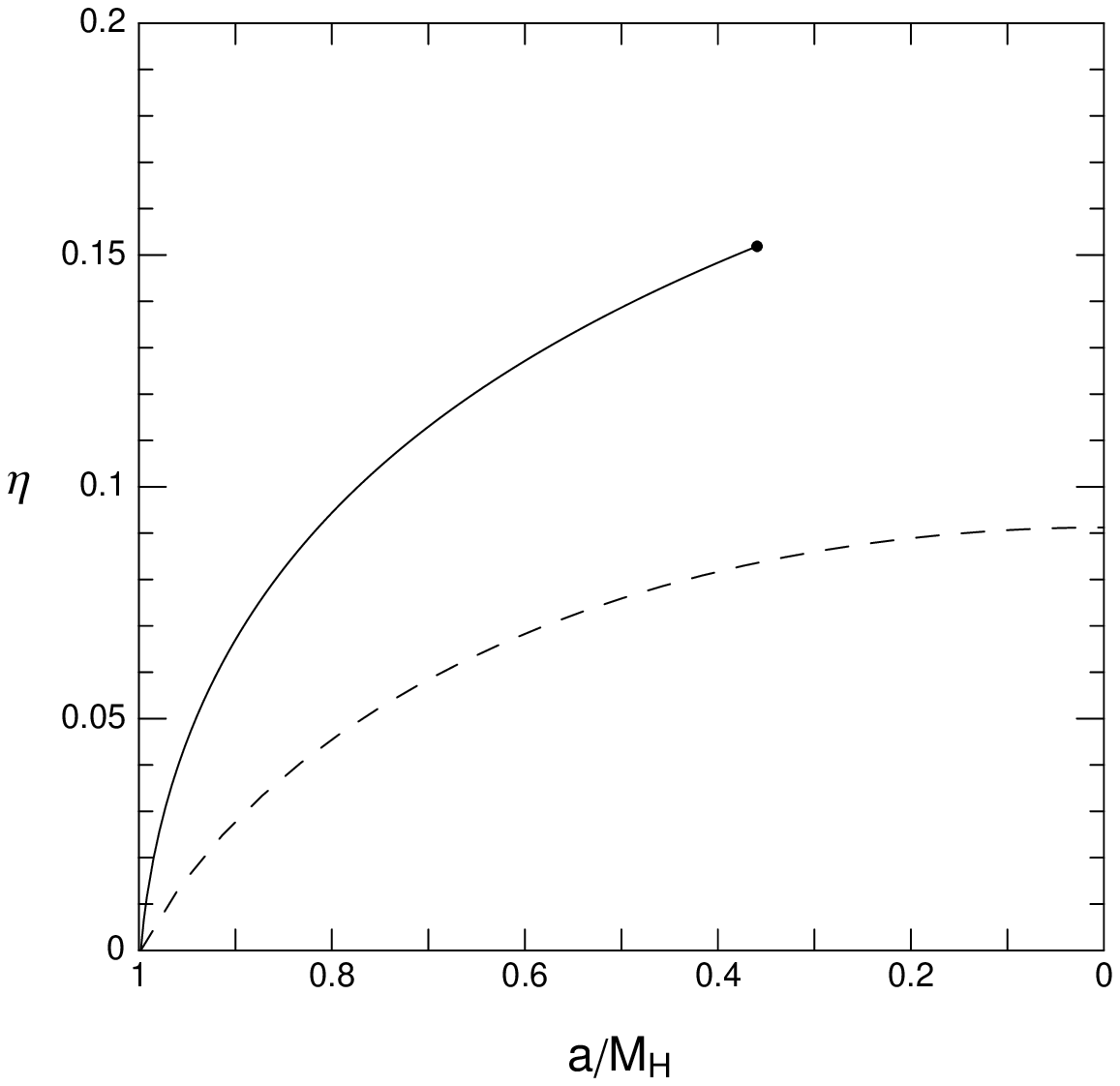]{The efficiency in extracting energy from a Kerr black hole
as the black hole is spun down. The efficiency is defined by $\eta = \Delta E/M_H$,
where $M_H$ is the mass of the black hole at its initial state with $a/M_H = 0.998$
(the maximum value of $a/M_H$ that an astrophysical black hole can have). 
(So the left ends
of the curves are at $a/M_H = 0.998$, not $a/M_H = 1$. Note that in the figure 
$a/M_H$ decreases from left to right.) The solid curve represents
the efficiency in extracting energy from a Kerr black hole through a thin Keplerian 
disk, which ends at $a/M_H = 0.36$ since then the transfer of energy and angular 
momentum from the black hole to the disk stops. With this mechanism, up to $\approx 
15\%$ of the initial mass of the black hole can be extracted. The dashed curve represents
the efficiency of the Blandford-Znajek mechanism, which ends at $a/M_H = 0$. With the
Blandford-Znajek mechanism, up to $\approx 
9\%$ of the initial mass of the black hole can be extracted. 
\label{figure2}}

\end{document}